\documentclass[final,5p,twocolumn,10pt,sort&compress]{elsarticle}

\usepackage[english]{babel}
\usepackage{enumerate}  
\usepackage[shortlabels]{enumitem} 
\usepackage{amsmath, amssymb, amsthm}
\usepackage{empheq} 
\usepackage{graphicx}
\usepackage{appendix}
\usepackage{comment}
\usepackage{subcaption}
\usepackage{booktabs,dcolumn,caption}
\usepackage{tabularx}
\usepackage{multicol} 
\usepackage{balance}
\usepackage{float}
\usepackage{xurl}
\usepackage{booktabs}   % nice rules
\usepackage{siunitx}    % for \num and scientific notation (optional)

\usepackage[dvipsnames]{xcolor}
\PassOptionsToPackage{
  colorlinks=true,
  linkcolor=MidnightBlue,
  citecolor=MidnightBlue,
  urlcolor=MidnightBlue
}{hyperref}
\usepackage{hyperref}

\usepackage[switch]{lineno}    % allow line numbers in 2-column mode
\usepackage{rotating}   % if you keep side-rotated version
\usepackage{adjustbox}  % optional: shrink to text-width
\newcommand{\sym}[1]{\textsuperscript{#1}}
\makeatletter
\def\ps@pprintTitle{%
 \let\@oddhead\@empty 
 \let\@evenhead\@empty
 \def\@oddfoot{}%
 \let\@evenfoot\@oddfoot}
 \makeatother

\usepackage{titlesec}

\titleformat{\section}
  {\bfseries\Large} % Large and bold for sections
  {\thesection}{1em}{} % Adjust spacing

\titleformat{\subsection}
  {\bfseries\normalsize} % Normal size but bold for subsections
  {\thesubsection}{1em}{} % Adjust spacing

\begin{document}

\hypersetup{ % overwrite hyperlink settings
  colorlinks=true,
  linkcolor=MidnightBlue,
  citecolor=MidnightBlue,
  urlcolor=MidnightBlue
}

\begin{frontmatter}

\title{The Determinants of Judicial Promotion: Politics, Prestige, and Performance}

\author[1]{Ilya Davidson}
\author[2,3]{Sandro Claudio Lera\corref{cor1}}
\author[1,3]{Robert Mahari}

\cortext[cor1]{corresponding author: \texttt{slera@mit.edu}}
\address[1]{CodeX, The Stanford Center of Legal Informatics, Palo Alto, USA}
\address[2]{Institute of Risk Analysis, Prediction and Management, Southern University of Science and Technology, Shenzhen, China}
\address[3]{Connection Science, Massachusetts Institute of Technology, Cambridge, USA}

% Abstract and keywords. 
%%%%%%%%%%%%%%%%%%%%%%%%%%%%%%%%%%%%%%%%%%%%%%%%%%%%%%%%%%%%%%%%%%%%%%%%%%%%%%%%%%%
\begin{abstract}
Judicial promotions shape the composition of higher courts, yet their determinants remain poorly understood. This paper examines promotion from U.S. District Courts to Courts of Appeals using a discrete-time hazard framework that models annual promotion probability. Using a judge-year panel covering over 36,000 observations from 1930 to present, we incorporate career timing, political alignment, elite credentials, and judicial performance measures.
Promotion probabilities follow a life-cycle pattern and are strongly influenced by political alignment between judges and presidents ($\beta$ = 2.12, p < 0.001). Elite credentials and productivity increase promotion likelihood, while higher reversal rates reduce it. Citation network centrality exhibits a meaningful association ($\beta$ = 0.230, p = 0.025) that operates independently of elite credentials. Promotion outcomes reflect a dynamic process shaped by timing, politics, elite networks, and performance signals, with political considerations dominating but not eclipsing judicial behavior.
\end{abstract}

\end{frontmatter}

%%%%%%%%%%%%%%%%%%%%%%%%%%%%%%%%%%%%%%%%%%%%%%%%%%%%%%%%%%%%%%%%%%%%%%%%%%%%%%%%%%%
%%%%%%%%%%%%%%%%%%%%%%%%%%%%%%%%%%%%%%%%%%%%%%%%%%%%%%%%%%%%%%%%%%%%%%%%%%%%%%%%%%%
\section*{Introduction}
\label{sec:intro}
%%%%%%%%%%%%%%%%%%%%%%%%%%%%%%%%%%%%%%%%%%%%%%%%%%%%%%%%%%%%%%%%%%%%%%%%%%%%%%%%%%%
%%%%%%%%%%%%%%%%%%%%%%%%%%%%%%%%%%%%%%%%%%%%%%%%%%%%%%%%%%%%%%%%%%%%%%%%%%%%%%%%%%%

Judicial promotions shape the composition of higher courts and, by extension, the development of the law. This is particularly consequential in common-law systems, where judicial opinions do not merely resolve individual disputes but also establish binding precedent that governs future cases \cite{holmes1881}. The higher the court, the broader the binding reach of its decisions. Because opinions issued by higher courts constrain the rulings of lower courts, promotion decisions determine which judges gain the authority to shape the law beyond individual cases \cite{cross2007decision}. Promotions therefore matter not only for judicial careers, but for legal development itself: elevating particular judges elevates particular interpretive approaches, doctrinal priorities, and styles of reasoning into positions of greater authority. Understanding how judges are selected for promotion is thus central to understanding how the judiciary evolves over time.  

Judicial promotion is often framed in meritocratic terms, analogous to advancement in other professional careers: a reward for legal skill \cite{davidow1981beyond, mojon2025data}, diligence \cite{EpsteinKnight2000JudicialPolitics}, and professional reputation \cite{landes1998judicial}. This view resonates with longstanding norms of judicial independence and the expectation that rulings be insulated from overt political influence \cite{holmes1881, Bickel1962LeastDangerousBranch}. At the same time, judicial appointments occur within explicitly political institutions, raising the possibility that merit-based and partisan considerations coexist in shaping promotion outcomes \cite{BonicaSen2021EstimatingIdeology}. Judges operate within a labor market of ideas and prestige, where incentives such as citation, reputation, and the prospect of elevation can influence performance \cite{, Baum2006JudgesAudiences, landes1980legal, shepherd2009influence}. In this context, judges may “audition” for higher office, consciously or subconsciously tailoring their behavior in anticipation of promotion. 

Promotion occupies a distinctive role within this incentive structure. Federal judges enjoy extraordinary job security: they rarely get fired \cite{USConstitutionArticleIII}, face limited oversight and operate under only loose constraints imposed by workload and appellate review. Furthermore, promotions are rare and judges serve for life \cite{PosnerEpsteinLandes2013BehaviorFederalJudges}. Each appointment therefore represents a durable investment for the appointing political coalition. From the perspective of presidents and senators, elevating a judge is not merely an assessment of professional competence, but an opportunity to shape the ideological and institutional trajectory of the courts for decades \cite{MoraskiShipan1999PoliticsNominations}. In this setting, considerations such as ideological alignment, perceived institutional loyalty, and demographic representation may sit alongside, or even outweigh, traditional measures of merit. Judicial promotion thus emerges not as a purely technocratic evaluation of legal excellence, but as the outcome of strategic selection within a low-incentive environment, where promotion serves simultaneously as a reward for judges and as a tool of political influence. 

Prior research emphasizes political alignment and strategic signaling rather than purely legal merit \cite{moraski1999politics, binder2009advice}. Traditional measures of competence - such as reversal rates, citation counts or productivity - have been found not to meaningfully predict promotion, even though judges who are promoted remain highly competent in their subsequent appellate careers \cite{ChoiGulatiPosner2015RoleCompetencePromotions}. This suggests that competence alone is neither a necessary nor a sufficient condition for advancement. Complementing this view, evidence indicates that appellate judges with realistic prospects of Supreme Court nomination systematically alter their behavior during vacancy periods: they rule more frequently in favor of the federal government, align more closely with presidential preferences, and increase visible judicial activity \cite{black2016courting}. Judges without plausible promotion prospects exhibit no such changes, indicating that these patterns reflect strategic signaling rather than fixed ideological differences.

Taken together, previous work portrays judicial advancement as a politically mediated process shaped by signaling and institutional incentives. However, it has focused primarily on how judges adjust their behavior in anticipation of promotion, often within narrow vacancy windows, rather than on which behaviors are ultimately rewarded. Our approach is complementary but distinct. Rather than asking how judges behave when promotion is possible, we ask which judicial behaviors increase the probability of promotion over time. By modeling promotion as a dynamic process, we shift attention from judicial signaling to the revealed selection criteria of political actors, identifying the characteristics that presidents and senators systematically reward when making promotion decisions.

Progress on this question has been constrained by data limitations \cite{pah2020build}. Most empirical studies on judicial promotion rely on small samples, narrow time frames, or have focused on higher courts, like the U.S. Supreme Court, where data is more readily available \cite{carp_stidham_manning_federal_judges, ruger2004supreme}. Yet the vast majority of legal disputes are resolved in the lower courts. In the federal system, district courts handle far more cases than appellate or supreme courts and play a critical role in interpreting and applying law in routine and consequential contexts. Despite their institutional importance, systematic quantitative research on judicial behavior and promotion at the district level has been limited by fragmented data and the absence of comprehensive longitudinal records.
Recent advances in data digitization now make it possible to overcome these constraints. In this study, we construct a longitudinal panel of federal judges by combining biographical and appointment data from the Federal Judicial Center with case-level data from the Caselaw Access Project (CAP). We apply natural language processing (NLP) techniques to obtain information on 360,087 civil U.S. District Court cases dating back to 1880 and link them to the judges who authored the corresponding opinions. To assess subsequent appellate outcomes, we incorporate 90,445 federal appellate decisions obtained from CourtListener. These appellate rulings are matched to the underlying district court opinions, allowing us to identify whether and how often district court decisions were reversed.

We follow 2,588 judges throughout their district court careers, up to the point of elevation where applicable. The linked dataset enables the construction of career-level measures—including reversal rates, opinion length, citation centrality \cite{Fowler2007NetworkAnalysisLaw}, and productivity—allowing us to examine how professional performance, political alignment, institutional positioning, and background characteristics shape the timing and likelihood of promotion within the federal judiciary.

To study these dynamics, we employ a discrete-time hazard model\cite{Allison1982DiscreteTimeMethods, BoxSteffensmeierJones2004EventHistory} that treats judicial promotion as a time-to-event process. Rather than asking which judges are eventually promoted, this framework estimates the probability that a judge is promoted in a given year, conditional on having remained unpromoted and eligible up to that point. This approach allows promotion determinants to vary over the course of a judge’s career, properly accounts for right-censoring among judges who are never promoted, and avoids the selection bias inherent in models that condition on promotion ex post. Prior empirical work in the judicial literature has relied almost exclusively on static or pooled specifications that collapse career trajectories into a single observation, thereby obscuring how the influence of competence, ideology, and institutional alignment evolves over time. By introducing a dynamic hazard-based framework, this paper offers a more realistic account of judicial promotion.

%%%%%%%%%%%%%%%%%%%%%%%%%%%%%%%%%%%%%%%%%%%%%%%%%%%%%%%%%%%%%%%%%%%%%%%%%%%%%%%%%%%
%%%%%%%%%%%%%%%%%%%%%%%%%%%%%%%%%%%%%%%%%%%%%%%%%%%%%%%%%%%%%%%%%%%%%%%%%%%%%%%%%%%
\section*{Data and Methods}
\label{sec:methods}
%%%%%%%%%%%%%%%%%%%%%%%%%%%%%%%%%%%%%%%%%%%%%%%%%%%%%%%%%%%%%%%%%%%%%%%%%%%%%%%%%%%
%%%%%%%%%%%%%%%%%%%%%%%%%%%%%%%%%%%%%%%%%%%%%%%%%%%%%%%%%%%%%%%%%%%%%%%%%%%%%%%%%%%

We construct a judge–year panel of federal district court judges observed annually from appointment until promotion to the appellate bench, retirement, death, or the end of the sample period. The final dataset contains 36,194 judge–years drawn from 2,588 judges.

The outcome of interest is promotion to a federal appellate court. We code a binary indicator equal to one in the year a judge is promoted and zero otherwise. Judges exit the observation period following promotion. Promotion events are rare, occurring in approximately 0.64 percent of judge–years.

We model promotion using a discrete-time hazard framework estimated via logistic regression. This approach captures the conditional probability that a judge is promoted in year $t$, given that they have not been promoted previously, and allows promotion hazard to depend on both time-varying and time-invariant characteristics. We control for judicial age and years of service on the district court to account for life-cycle patterns in promotion. We further include decade-of-appointment fixed effects to absorb cohort-specific differences in promotion opportunities and circuit fixed effects to control for persistent institutional variation across jurisdictions. Standard errors are clustered at the judge level to account for serial correlation within judicial careers.

We now describe the key independent variables; Table~\ref{tab:descriptive_stats} provides summary statistics.
Political alignment is measured using an indicator for copartisanship, which equals one if the president who appointed the judge shares the same party affiliation as the sitting president in year \textit{t}, and zero otherwise. We also include an indicator for the appointing president's party, coded as one if the judge was appointed by a Republican president and zero if appointed by a Democrat. Promotion requires both presidential nomination and Senate confirmation; our alignment measures capture this joint selection process (see \ref{app:politicalalignment} for details). 

We measure credential tier using indicators for attendance at a top-ranked law school and completion of an appellate-level judicial clerkship. A law school is classified as elite if it ranked among the top five institutions according to U.S. News \& World Report in 1987 (the earliest year rankings are available):  Harvard, Yale, Columbia, Michigan or Stanford. A clerkship is considered elite if the judge served as a law clerk for a Justice of the U.S. Supreme Court or for a judge on a U.S. Court of Appeals. 

We incorporate multiple measures of judicial performance and decision behavior using the published written opinions authored by district judges. Judicial opinions are formal written decisions resolving cases and are publicly available through federal reporting systems. These opinions contain legal reasoning and citations to prior cases, allowing us to observe both writing characteristics and citation patterns.

All performance and behavioral measures are constructed cumulatively using only information available up to each judge-year. For a judge in year $t$, we calculate measures based exclusively on cases decided from their appointment through the end of year $t-1$, ensuring no look-ahead bias. This creates time-varying measures that evolve as judges accumulate experience and build track records. All measures are standardized to have mean zero and unit variance to facilitate comparison of effect sizes. Technical construction details, including empirical Bayes smoothing procedures and exact formulas, are provided in \ref{app:judicialperformance}. We now describe each measure.

First, we measure appellate reversal rates. When district court decisions are appealed to circuit courts, appellate judges review the lower court's legal reasoning and may affirm (uphold) or reverse (overturn) the decision. We classify any non-affirmative outcome—including reversals, vacations, or remands—as a reversal. Reversal rates provide a direct signal of legal accuracy: judges who are reversed more frequently deviate more often from appellate court precedent or commit legal errors. 

Second, we measure judicial influence through citation network centrality. Legal reasoning operates as a citation system: judges establish the validity of their rulings by citing prior cases as precedent \cite{landes2013legal}. This citation structure creates a network where some opinions become foundational authorities while others remain peripheral \cite{fowler2007network}. Network-based measures of citation patterns reveal the underlying structure of legal doctrine and the relative influence of judicial decisions within it, as recent work shows citation dynamics in law share similar statistical properties with scientific and patent citations \cite{kojaku2025communitycentricmodelingcitationdynamics}. 
We construct this citation network from our dataset of cases, where each case is a node and each citation is a directed edge. When Case A cites Case B in its legal reasoning, this creates an edge from A to B, indicating B influenced A. Using our dataset of cases we construct this citation network. We measure each case's centrality using PageRank, which assigns higher centrality to cases cited by other central (influential) cases. For each judge, we calculate the average PageRank centrality of all authored opinions. This network-based approach distinguishes between judicial productivity and prominence within the legal network, recognizing that citations from landmark precedents carry more weight than citations from obscure decisions. We winsorize at the 1st and 99th percentiles to reduce outlier influence. Results with raw citation counts can be found in  \ref{app:judicialperformance}.

Third, we measure publication intensity as the cumulative number of authored, published opinions divided by years of service, yielding an average publication rate per year. We log-transform publication rates to account for right-skewness in the distribution. Unlike citation centrality, publication intensity captures output intensity and visibility rather than downstream influence. This distinction allows us to separate productivity from network prominence. By normalizing cumulative publications by years served, the measure avoids mechanically increasing with tenure and instead reflects differences in publication intensity conditional on career length.

Fourth, we include average opinion length, measured as the mean of log word counts. This variable proxies for writing style and effort allocation, capturing whether judges produce shorter, efficiency-oriented opinions or longer, more elaborative reasoning.

To assess judicial behavior toward powerful litigants, we measure decision patterns in cases involving the federal government and large corporations.
In civil cases where the United States (or a federal agency) appears as either plaintiff or defendant, we measure how often the judge rules in favor of the government. We code case outcomes as pro-government (government wins on dispositive motion, receives favorable judgment, or prevails at trial) or anti-government (government loses, motion denied, or unfavorable judgment). The government support rate is the proportion of pro-government outcomes among all resolved cases involving the government.
For cases involving large corporations as defendants—typically in disputes with employees, consumers, or regulators—we similarly code outcomes as pro-corporate (e.g., dismissal of claims, summary judgment for the defendant, or defense verdict) or anti-corporate (e.g., denial of dismissal, plaintiff summary judgment, or plaintiff verdict). The company favorism rate is the proportion of pro-corporate outcomes among such cases. Together, these measures capture systematic directional tendencies in rulings involving governmental and business interests.

We next allow the association between judicial characteristics and the annual promotion hazard to vary by presidential party by interacting key performance and behavioral measures with an indicator for Republican administrations. This specification tests whether Democratic and Republican presidents systematically differ in how they weight judicial signals when making promotion decisions.

%%%%%%%%%%%%%%%%%%%%%%%%%%%%%%%%%%%%%%%%%%%%%%%%%%%%%%%%%%%%%%%%%%%%%%%%%%%%%%%%%%%
%%%%%%%%%%%%%%%%%%%%%%%%%%%%%%%%%%%%%%%%%%%%%%%%%%%%%%%%%%%%%%%%%%%%%%%%%%%%%%%%%%%
\section*{Results} 
\label{sec:results}
%%%%%%%%%%%%%%%%%%%%%%%%%%%%%%%%%%%%%%%%%%%%%%%%%%%%%%%%%%%%%%%%%%%%%%%%%%%%%%%%%%%
%%%%%%%%%%%%%%%%%%%%%%%%%%%%%%%%%%%%%%%%%%%%%%%%%%%%%%%%%%%%%%%%%%%%%%%%%%%%%%%%%%%
Our analysis proceeds in three parts. First, we document historical patterns in judicial promotions, describing how rates of elevation vary by era, administration, and circuit. Second, we estimate the relative weight of political, professional, and institutional factors in predicting annual promotion propensities, conditional on continued eligibility for elevation. Third, we explore heterogeneity across presidential administrations, asking whether certain presidents favor competence over ideology, in shaping the appellate bench.

\subsection*{Descriptive statistics}
 Table~\ref{tab:descriptive_stats} reports descriptive statistics for the judge-year panel used in the hazard analysis. The sample consists of 36,194 judge-years drawn from 2,588 federal judges, during which 232 promotions occur, corresponding to an annual promotion probability of approximately 0.64 percent. Promotions are thus rare events, underscoring the appropriateness of a discrete-time hazard framework.

The distribution of judge-years across tenure stages varies substantially within the sample. Approximately 69 percent of all judge-years occur within the first decade of tenure, while only about 11 percent occur after sixteen years on the bench, consistent with the expectation that promotion opportunities are concentrated early in judicial careers. Judges are observed under Democratic and Republican administrations in roughly equal proportions, and just over half of judge-years involve a copartisan president.

Elite credentials are relatively uncommon. Approximately one-fifth of judge-years involve judges with elite law school backgrounds, and just over ten percent involve judges with appellate-level clerkships (Supreme Court or Courts of Appeals). 

The sample remains demographically skewed, with men accounting for over eighty percent of judge-years and White judges comprising approximately eighty-five percent. These distributions provide important context for interpreting subsequent regression estimates, which assess promotion probabilities conditional on this underlying composition. These numbers can be found in \ref{tab:composition}.

\begin{table*}[htbp]
\centering
\caption{Descriptive Statistics and Panel Overview}
\label{tab:descriptive_stats}
\small
\setlength{\tabcolsep}{6pt}
\begin{tabular}{lrrrrrr}
\toprule
\multicolumn{7}{l}{\textit{Panel A: Panel Overview}} \\
\midrule
Judge--years & \multicolumn{6}{r}{36,194} \\
Judges & \multicolumn{6}{r}{2,588} \\
Promotions (events) & \multicolumn{6}{r}{232} \\
Annual promotion rate & \multicolumn{6}{r}{0.0064} \\
\midrule
\multicolumn{7}{l}{\textit{Panel B: Distribution of Covariates (Judge--Year Level)}} \\
\midrule
 & Mean & Min & 25th pctl. & Median & 75th pctl. & Max \\
\midrule
Age & 57.32 &  33.00 &  52.00 & 58.00 & 63.00 & 79.00 \\
Copartisan president & 0.536 & 0.000 & 0.000 & 1.000 & 1.000 & 1.000 \\
Elite law school & 0.200 & 0.000 & 0.000 & 0.000 & 0.000 & 1.000 \\
Elite clerkship & 0.118 & 0.000 & 0.000 & 0.000 & 0.000 & 1.000 \\
Reversal rate (EB, cumulative) & 0.000 & -3.079 & -0.664 & -0.018 & 0.379 & 4.441 \\
Citation Centrality & -0.019 & -1.760 & 0.012 & 0.073 & 0.316 & 2.452 \\
Publication intensity (log, cumulative) & 0.000 & -1.569 & -0.737 & 0.088 & 0.720 & 3.298 \\
Opinion length (average, log) & 0.000 & -7.301 & -0.457 & 0.021 & 0.496 & 5.769 \\
Government support rate (EB) & 0.000 & -4.225 & -0.567 & 0.032 & 0.726 & 3.161 \\
Company favorism rate (EB) & 0.000 & -3.065 & -0.642 & -0.124 & 0.545 & 4.782 \\
\bottomrule
\end{tabular}

\begin{flushleft}
\footnotesize
Notes: The unit of observation is the judge--year. Promotion is defined as elevation from a U.S. district court to a U.S. court of appeals. Performance and behavioral measures are standardized to have mean zero and unit variance. Empirical Bayes (EB) smoothing is applied to rate variables constructed from cumulative case counts. Citation centrality is winsorized at the 1st and 99th percentiles to reduce the influence of extreme outliers; the non-zero mean (-0.019) reflects this adjustment.
\end{flushleft}
\end{table*}

\subsection*{Baseline determinations of judicial promotion}
Before turning to the main covariates of interest, we first establish that the basic structure of judicial promotion probabilities in our data aligns with well-documented institutional and career patterns. We begin by estimating a baseline discrete-time hazard specification that captures the role of career timing and institutional context in promotions from the U.S. District Courts to the Courts of Appeals.

As reported in Table~\ref{tab:promotion_models}, this specification includes controls for judicial age, a flexible set of tenure indicators, cohort fixed effects based on era of appointment, circuit fixed effects, and baseline demographic characteristics. These covariates jointly characterize the background promotion environment against which the effects of political alignment, elite credentials, and judicial performance are subsequently evaluated.

Promotion probabilities decline sharply with age. The linear age term is negative and highly statistically significant ($\beta$ = -0.133, p $<$ 0.001), indicating that promotion likelihood decreases at an accelerating rate as judges age. 

As shown in Table~\ref{tab:promotion_models}, promotion likelihood increases with judicial tenure relative to the 0--2 year baseline. Judges with three to five years of service are significantly more likely to be promoted ($\beta$ = 0.936, p < 0.01), and this elevated hazard persists across subsequent tenure categories. Promotion probabilities continue to rise through mid-career, reaching their highest levels among judges with 20 or more years on the district bench ($\beta$ = 2.127, p < 0.01). The pattern is broadly increasing and then flattens at longer tenures, consistent with a life-cycle dynamic in which promotion opportunities expand with accumulated experience before leveling off.

Appointment cohort significantly predicts promotion probability across all models. Using 1930s appointees as the reference category, judges appointed in later decades face systematically lower promotion probabilities. This temporal decline becomes statistically significant beginning with 1990s appointees ($\beta$ = -1.61, p < 0.001) and is especially pronounced for judges appointed in the 2000s ($\beta$ = -2.67, p < 0.001), 2010s ($\beta$ = -2.13, p < 0.001), and 2020s ($\beta$ = -1.94, p < 0.001). These cohort effects likely reflect structural changes in the federal judiciary, including declining appellate court vacancy creation, increased judicial tenure and delayed turnover, and changing political norms surrounding promotion and Senate confirmation.

By contrast, circuit-level differences are comparatively modest. While a small number of circuits exhibit statistically significant coefficients—most notably the Second, Fourth, and Sixth Circuits—the majority of circuit fixed effects are statistically indistinguishable from zero. Overall, these results suggest that once cohort and career timing are accounted for, promotion patterns are broadly similar across geographic jurisdictions.

These results are robust to alternative specifications of the baseline hazard and to controls for tenure and age separately. Even among judges who remain well within conventional age limits for promotion, additional years on the bench are associated with lower promotion probabilities. 

Taken together, these covariates demonstrate that judicial promotion is partially structured by institutional timing and cohort effects. Age and tenure exert strong, predictable influences on promotion probabilities, while circuit-level variation plays a secondary role. These results provide a clear baseline against which the effects of political alignment, elite credentials, and performance-based measures are evaluated in subsequent models.

\subsection*{Political Alignment and Elite Credentials}

We now discuss political alignment and elite credentials. The results indicate that copartisanship is the single strongest predictor of promotion in the model. Judges serving under a copartisan president are substantially more likely to be promoted ($\beta$ = 2.117, p < 0.001), corresponding to an odds ratio of approximately 8.05. This finding is consistent with prior research demonstrating that same-party appointment is the strongest predictor of judicial elevation \cite{savchak2006taking} and with broader evidence of increasing partisan alignment effects in federal judicial decision-making \cite{cohen2024judges, whittington2018partisanship}.

Elite credentials are also positively associated with promotion outcomes. Judges with elite law school backgrounds exhibit significantly higher promotion probabilities ($LS_p$: $\beta$ = 0.322, p = 0.005), as do judges who completed prestigious clerkships ($C_p$: $\beta$ = 0.566, p = 0.008). While prior research has documented the importance of elite credentials in initial judicial appointments and the prestige associated with federal clerkships \cite{fogel2023law}, our findings extend this literature by demonstrating that these credentials also predict advancement within the federal judiciary. These results are consistent with theories emphasizing the role of elite signaling and professional networks in judicial careers \cite{black2016courting, badas2025judicial}.

\subsection*{Performance and decision behavior}
We now discuss measures of judicial performance and decision behavior. Reversal rates are a notable promotion indicator. Judges who are reversed more frequently by higher courts face a significantly lower hazard of promotion, even after controlling for political alignment, elite credentials, and career timing ($\beta$=-0.159, p=0.016). This finding indicates that selectors do attend to at least some performance signals, particularly those that provide direct and salient feedback from superior courts. This finding differs from existing literature, which does not report a statistically significant association between reversal rates and promotion \cite{ChoiGulatiPosner2015RoleCompetencePromotions}. When we restrict our sample to the time period examined in that study (2001-2012), we likewise fail to detect a significant effect. The difference appears to stem primarily from our expanded data coverage, which substantially increases the number of promotion events and thus statistical power. 

Citation network centrality exhibits another meaningful association with promotion ($\beta$ = 0.230, p = 0.025), indicating that judicial influence within the legal citation network operates as a performance-based signal alongside political and credential factors. A one-standard-deviation increase in centrality increases promotion odds by approximately 26\%, comparable to reversal rates and other performance measures. Without winsorization, the coefficient is substantially smaller ($\beta$ = 0.071, p = 0.006), suggesting that a small number of judges with extraordinarily influential landmark cases were not promoted and suppressed the average effect. Raw citation counts remain non-significant (not shown), confirming that citation volume alone does not predict advancement. This distinction between network centrality and raw citations indicates that selectors distinguish judges whose cases are cited by other influential decisions from those who simply accumulate high citation volumes.

Judicial workload also predicts promotion. Publication intensity is strongly and positively associated with promotion probability ($\beta$ = 0.361, p < 0.001), indicating that judges who produce more written opinions face substantially higher promotion hazards. This effect persists after conditioning on age, tenure, political alignment, and elite credentials, suggesting that visible effort and output—rather than doctrinal influence per se—are rewarded in the promotion process. Because publication volume is normalized by years of service (using the log average publication rate), this result cannot be attributed mechanically to longer careers; instead, it reflects differences in publication intensity conditional on tenure.

Taken together, the results reveal that judicial performance matters for promotion, though its influence remains secondary to political alignment and elite credentials. Citation network centrality ($\beta$ = 0.23), publication intensity ($\beta$ = 0.36), and reversal rates ($\beta$ = -0.16) each predict advancement, suggesting selectors attend to judicial influence, productivity, and legal accuracy—though political considerations dominate. In the next section, we examine whether performance signals matter differently depending on the president's party.

\subsection*{Party-conditional interaction effects}
If promotion decisions reflect partisan screening, the value of specific forms of decision behavior should depend on the partisan affiliation of the sitting president. To assess this possibility, we interact behavioral measures with an indicator for Republican presidencies, allowing their association with promotion to vary across administrations, which is shown in Table \ref{tab:party_interactions}. 

The association between government-aligned rulings and promotion differs by party. Under Democratic presidents (the baseline category), the government support rate is not significantly related to promotion ($\beta = -0.123$, p = 0.189). However, the interaction with a Republican presidency is positive and precisely estimated ($\beta = 0.414$, p = 0.002), implying a positive net association under Republican administrations ($\beta = 0.290$). Government-aligned decision behavior thus appears to be rewarded primarily under Republican presidents.

In contrast, corporate-aligned decision behavior shows no association with promotion under either partisan regime. Neither the Democratic effect ($\beta = 0.085$, p = 0.320) nor the Republican total effect ($\beta = -0.068$, p = 0.250) is statistically distinguishable from zero.

Reversal rates and citation centrality, while associated with promotion in the pooled model, do not exhibit meaningful party-conditional differences. The interaction term is small and statistically insignificant, suggesting that any performance penalties or rewards operate similarly across administrations.

\begin{table}[htbp]
\centering
\caption{Determinants of Judicial Promotion in a Discrete-Time Hazard Model}
\label{tab:promotion_models}
\small
\begin{tabular}{l
S[table-format=-1.3]}
\hline\hline
\multicolumn{2}{l}{\textit{Panel A: Career Timing and Demographics}} \\
\hline 

Age                 & -0.133\sym{***} \\
Age$^2$             & -0.006\sym{***} \\
Tenure: 3--5 yrs    &  0.936\sym{**}  \\
Tenure: 6--10 yrs   &  1.086\sym{**}  \\
Tenure: 11--15 yrs  &  1.529\sym{**}  \\
Tenure: 16--20 yrs  &  1.877\sym{**}  \\
Tenure: 20+ yrs     &  2.127\sym{**}  \\
Male                & -0.442\sym{**}  \\
\hline
\multicolumn{2}{l}{\textit{Panel B: Political Alignment and Elite Credentials}} \\
\hline
Copartisan          &  2.117\sym{***} \\
Elite Law School    &  0.322\sym{*}   \\
Elite Clerkship     &  0.566\sym{**}  \\
\hline
\multicolumn{2}{l}{\textit{Panel C: Performance and Decision Behavior}} \\
\hline
Reversal Rate            & -0.159\sym{*}   \\
Citation Centrality      &  0.230\sym{*}   \\
Publication Intensity    &  0.361\sym{***} \\
Opinion Length           &  0.091          \\
Government Support Rate  &  0.093          \\
Company Favorism Rate    & -0.002          \\
\hline
\multicolumn{2}{l}{\textit{Panel D: Baseline Hazard}} \\
\hline
Constant          & -6.590\sym{***} \\
                  & \multicolumn{1}{c}{(0.73)} \\
\hline
Era Fixed Effects     & \multicolumn{1}{c}{Yes} \\
Circuit Fixed Effects & \multicolumn{1}{c}{Yes} \\
Observations          & \multicolumn{1}{c}{36,194} \\
Judges                & \multicolumn{1}{c}{2,588}  \\
\hline\hline
\end{tabular}%
\begin{flushleft}
\footnotesize
Notes: Entries are log-odds coefficients from a discrete-time hazard model with judge-level clustered standard errors. Tenure 0--2 years is the reference category. White judges and Republican presidents are reference categories where applicable.
$\dagger p<0.10$, $^{*} p<0.05$, $^{**} p<0.01$, $^{***} p<0.001$.
\end{flushleft}
\end{table}

\subsection*{Mechanism: Endogeneity of Reversal Rates}
Our results show that judges with lower reversal rates are more likely to be promoted. However, this relationship admits two interpretations. First, reversal rates may reflect judicial quality: some judges make more legal errors and are therefore reversed more often. Under this interpretation, presidents observe reversal rates and use them as a signal of competence when selecting candidates for promotion. Second, reversal rates may reflect random variation in panel assignment: appellate cases are randomly assigned to three-judge panels, and some panels are systematically harsher than others. A district judge who happens to draw reversal-prone appellate panels will accumulate a higher reversal rate despite writing legally sound opinions. We call this "bad luck" rather than "bad judgment".

If presidents cannot distinguish these two sources of variation, judicial quality versus random panel assignment, then our baseline estimate conflates genuine quality screening with arbitrary outcomes.  The random assignment of district court appeals to appellate panels \cite{abrams2007luck, mahari2024earlycareercitationscapture} provides a natural experiment well-suited to an instrumental variables design. Because panel composition is random, variation in a district judge's reversal rate driven by the harshness of the panels they face is orthogonal to their underlying judicial quality, allowing us to isolate the causal effect of reversals on promotion.

We construct an instrumental variable that captures this random component. For each district judge $d$, we calculate the average reversal propensity of the appellate panels that reviewed their decisions. Specifically, for each appellate case involving district judge $d$, we compute the mean reversal rate of the appellate judges on that panel (excluding cases involving judge $d$ to avoid mechanical correlation). We then average these panel-level propensities across all appeals involving judge $d$. This instrument—which we call "panel harshness exposure"—measures the extent to which a district judge was randomly exposed to harsh versus lenient appellate review, independent of their own case quality or judicial skill. The instrument satisfies the standard IV conditions: it is relevant, as panel reversal propensity meaningfully predicts the district judge's own reversal rate; it is exogenous, as random panel assignment ensures independence from the district judge's underlying quality; and it satisfies the exclusion restriction, as a panel's general tendency to reverse plausibly affects promotion prospects only insofar as it shapes the district judge's reversal record, with no direct channel to promotion decisions.

Our results show that the first-stage relationship is moderate (F $\approx$ 7), indicating the instrument has meaningful but limited predictive power over observed reversal rates. However, this is also attributable in part to the reduced sample size of the IV analysis, which is restricted to judges with at least three appealed decisions and retains only 116 promotion events compared to 232 in the baseline specification, limiting statistical power. Results are reported in ~\ref{app:ivanalysis}. The IV estimate on reversal rates is positive and the control-function residual is negative, though neither reaches conventional significance levels (p = 0.29 and p = 0.26, respectively). This sign pattern suggests that presidents penalize reversals reflecting genuine judicial errors (the endogenous component) but not those driven by random panel assignment (the exogenous component). While the limited instrument strength prevents definitive causal conclusions, the analysis is consistent with reversal rates operating as a meaningful quality signal rather than mechanically penalizing judges for panel composition.

\subsection*{Mechanisms Behind Elite Clerkship Effects}
To better understand why elite clerkships are associated with higher promotion probabilities, we examine whether their effect reflects selection on unobserved ability or post-appointment mechanisms such as training, visibility, or professional networks.

We begin by analyzing early-career behavior among district judges. Focusing on the first five years after appointment, we regress several performance measures on elite credential status while controlling for demographic characteristics and cohort effects. Table~\ref{tab:early_career_behavior} shows that judges from top law schools exhibit higher early productivity (more publications), write longer opinions, and receive more citations, but do not differ systematically in reversal rates. Elite clerks similarly publish more early in their careers but show no differences in opinion length, citations, or reversals. This pattern suggests that elite credentials select for judges who are more productive and visible early on, but not necessarily more accurate in their legal reasoning. Importantly, while elite law school graduates write more central cases early in their careers, this does not fully explain their promotion advantage: controlling for centrality reduces the law school coefficient by only 19\% (from $\beta$ = 0.40 to $\beta$ = 0.32). This indicates that elite credentials operate primarily through reputational signaling rather than through demonstrated legal influence, even though centrality does independently predict promotion ($\beta$ = 0.230, p = 0.025).

To more formally test whether the clerkship premium evolves over time, we estimate interaction models between clerkship status and tenure bins. Table~\ref{tab:clerkship_tenure_interactions} shows the clerkship effect is largest immediately upon appointment ($\beta$ = 1.44, p < 0.001) but declines significantly during the subsequent decade (interactions at 3-5 years: $\beta$  = -1.59, p < 0.01; 6-10 years: $\beta$  = -1.32, p < 0.05). This suggests elite clerkships create a fast-track pathway for early promotion, with clerks either elevated rapidly or experiencing a delayed premium that reemerges at longer tenures. Column (2) reports a placebo test using elite law school credentials, which shows no comparable tenure variation, confirming that the temporal pattern is specific to clerkships rather than elite credentials generally.

Finally, we assess whether the clerkship or law school effect persists within an already highly selected pool of judges. Restricting the sample to judges with elite observable credentials at appointment, we re-estimate the promotion hazard model as shown in Table~\ref{tab:within_elite_promotion}. We see that elite clerkship or law school status remains a strong predictor of promotion even within this elite subsample. This finding suggests that the elite premium cannot be fully attributed to baseline observable prestige and is consistent with mechanisms operating beyond simple selection.

\begin{table*}[!t]
\centering
\caption{Party Heterogeneity in Promotion Determinants}
\label{tab:party_interactions}
\begin{tabular}{l
    S[table-format=-1.3]
    S[table-format=-1.3]
    S[table-format=-1.3]}
\hline\hline
& \multicolumn{1}{c}{Democrats} & \multicolumn{1}{c}{Republicans} & \multicolumn{1}{c}{Difference} \\
& \multicolumn{1}{c}{(1)} & \multicolumn{1}{c}{(2)} & \multicolumn{1}{c}{(2)-(1)} \\
\hline
Reversal Rate             & -0.103 & -0.196\sym{**} & -0.093 \\
                          & {(0.105)} & {(0.085)} & {(0.134)} \\
Citation Centrality       & 0.209 & 0.234\sym{*} & 0.024 \\
                          & {(0.141)} & {(0.127)} & {(0.168)} \\
Government Support Rate   & -0.123 & 0.290\sym{***} & 0.414\sym{***} \\
                          & {(0.094)} & {(0.091)} & {(0.132)} \\
Company Favorism Rate     & 0.085 & -0.068 & -0.153 \\
                          & {(0.092)} & {(0.092)} & {(0.129)} \\
\hline\hline
\multicolumn{4}{l}{\footnotesize Estimated from model with party interactions.} \\
\multicolumn{4}{l}{\footnotesize All controls from Table 2 included.} \\
\multicolumn{4}{l}{\footnotesize $\sym{*}$ $p<0.10$, $\sym{**}$ $p<0.05$, $\sym{***}$ $p<0.01$}
\end{tabular}
\end{table*}
%%%%%%%%%%%%%%%%%%%%%%%%%%%%%%%%%%%%%%%%%%%%%%%%%%%%%%%%%%%%%%%%%%%%%%%%%%%%%%%%%%%
%%%%%%%%%%%%%%%%%%%%%%%%%%%%%%%%%%%%%%%%%%%%%%%%%%%%%%%%%%%%%%%%%%%%%%%%%%%%%%%%%%%
\section*{Discussion} 
%%%%%%%%%%%%%%%%%%%%%%%%%%%%%%%%%%%%%%%%%%%%%%%%%%%%%%%%%%%%%%%%%%%%%%%%%%%%%%%%%%%
%%%%%%%%%%%%%%%%%%%%%%%%%%%%%%%%%%%%%%%%%%%%%%%%%%%%%%%%%%%%%%%%%%%%%%%%%%%%%%%%%%%

This paper examines judicial promotion as a dynamic process, modeling the year-by-year probability that a judge is promoted conditional on continued eligibility. By adopting a discrete-time hazard framework, the analysis moves beyond pooled promotion outcomes and captures how promotion hazard evolves over a judge’s career as credentials accumulate, performance signals emerge, and political contexts change. The results reveal that judicial promotion is neither purely meritocratic nor purely ideological. Instead, it reflects shifting trade-offs between competence, visibility, and political alignment that vary systematically with the party controlling the presidency.

\subsection*{Political Alignment and Partisan Asymmetry}
A central finding concerns the dominant role of political alignment, particularly copartisanship between a judge and the sitting president. Across all specifications, copartisanship substantially increases the probability of promotion in a given year. However, the magnitude of the copartisanship coefficient may be misleading if interpreted as a simple linear predictor. Rather than functioning as one factor among many that presidents weigh continuously, copartisanship appears to operate more as a gating criterion or threshold condition: judges outside the president's partisan coalition face near-zero promotion probability regardless of their qualifications, while those within the coalition are then evaluated on performance and credentials. 

\subsection*{Career Timing and the Window of Promotion}

The analysis also highlights the importance of institutional timing. Promotion probabilities follow a pronounced life-cycle pattern, increasing during the early years of district court service, peaking after several years on the bench, and declining thereafter. Age exhibits a similar nonlinear relationship, with promotion hazard falling sharply at older ages. Together, these patterns support a “window of promotion” logic: judges face a limited period during which advancement is plausible, after which promotion prospects diminish regardless of accumulated experience.

These dynamics align with institutional realities of the federal judiciary. Promotion opportunities are episodic and contingent on vacancies, political alignment, and statutory or informal age norms. Judges who are not promoted within a certain window may age out of consideration, lose political salience, or be overtaken by newer cohorts who better match the priorities of current appointing authorities. The declining hazard also underscores the value of modeling promotion dynamically. Cross-sectional approaches that treat promotion as a static outcome obscure the fact that promotion prospects deteriorate over time even for otherwise qualified judges.

\subsection*{Demographics and Representational Considerations}

Demographic characteristics also play a systematic role in promotion outcomes. Conditional on credentials, behavior, and political alignment, male judges are less likely to be promoted than female judges. Similarly, Hispanic and Asian judges exhibit higher promotion hazards relative to White judges, while no consistent differences are observed for Black judges.

These patterns should be interpreted in light of the composition of the hazard set. Women and non-White judges remain substantially underrepresented among U.S. district court judges, reflecting earlier stages of selection into the federal judiciary. The estimated demographic effects therefore do not describe promotion prospects in the population at large, but rather conditional promotion probabilities among judges who have already been selected into the district courts. Within this highly selective and predominantly White, male pool, female judges and Hispanic and Asian judges exhibit higher promotion hazards. One plausible interpretation is that, once appointed to the district bench, these judges face comparatively stronger promotion prospects, potentially reflecting representational considerations or heightened scrutiny and selection at earlier stages of the judicial pipeline. Importantly, these results do not imply that demographic characteristics mechanically drive promotion, but rather that demographic differences persist even after conditioning on credentials, behavior, and political alignment.

\subsection*{Prestige and the Limits of Conventional Quality Measures}

The results draw a sharp contrast between different signals of judicial quality. Traditional prestige indicators—elite law school attendance and prestigious clerkships—are consistently associated with higher promotion probabilities. These credentials likely signal not only legal ability but also social capital, professional networks, and cultural compatibility with the upper echelons of the judiciary. Their persistence across specifications supports theories of elite reproduction and network-based advancement within professional hierarchies.

By contrast, ABA ratings perform poorly as predictors of promotion. ABA ratings do not significantly predict promotion in any specification. Once political alignment, elite credentials, and performance measures are included, the estimated association between ABA evaluations and promotion is statistically indistinguishable from zero. This pattern suggests that contemporary promotion decisions may rely less on formal merit assessments and more on political and reputational factors.

\subsection*{Performance Signals and the Limits of Legal Influence}

A key contribution of the paper is its finding that appellate reversal rates matter for promotion, and that they matter negatively. Judges who are reversed more frequently face lower promotion hazards, consistent with reputational models in which reversals signal legal error, lack of care, or ideological extremity. This result contrasts with prior work that found weak or null effects of reversals, suggesting that a dynamic framework and longer time horizon reveal performance signals that static models miss.

Citation network centrality exhibits a meaningful association with promotion, indicating that judicial influence within the legal citation network operates as a performance-based signal alongside political and credential factors. To ensure robustness to extreme outliers common in citation networks, we winsorize centrality at the 1st and 99th percentiles; without winsorization, the coefficient is substantially smaller, suggesting a small number of judges with extraordinarily influential landmark cases were not promoted and suppressed the average effect. While the magnitude remains smaller than copartisanship, it is comparable to reversal rates and other performance measures, indicating that selectors attend to judicial influence as measured by network position within the citation graph.

We explored whether centrality mediates the promotional advantage of elite credentials. While judges from top law schools write more central cases early in their careers (Table \ref{tab:early_career_behavior}, this does not fully explain their promotion advantage: controlling for centrality reduces the law school coefficient by only 19\%. Elite clerkships show no relationship with centrality, yet strongly predict promotion. Interactions between centrality and elite status are uniformly null, indicating that the promotional value of network prominence does not vary by credential background.

These patterns reveal that citation network centrality operates as an independent pathway to promotion rather than as a mechanism through which elite credentials confer advantages. Presidents and senators distinguish between raw citation volume—which does not predict promotion—and network position, rewarding judges whose cases are cited by other influential decisions. However, this effect emerges only among typical judges; those with extraordinarily central landmark cases face suppressed promotion probabilities, likely because such cases are rare and idiosyncratic rather than signaling systematic judicial quality. 

An instrumental variable analysis using leave-out panel harshness as an instrument for reversal rates yields a positive and insignificant estimate on the instrumented reversal rate, providing no evidence that judges are penalized for bad luck. The weak first stage limits causal inference, but the sign pattern reinforces rather than undermines the interpretation of reversal rates as a genuine performance signal. Stronger identification remains a direction for future work. 

\subsection*{Pro-Government Behavior and Career Incentives}

Judicial behavior toward the federal government in civil cases emerges as a substantively important, though nuanced, predictor of promotion. Judges who rule more frequently in favor of the government tend to face higher promotion hazards, particularly under Republican administrations. Pro-government rulings may thus signal institutional loyalty, ideological compatibility, or a restrained approach to judicial review.

At the same time, the magnitude of these effects is sensitive to specification, as the estimated coefficient on government support is modest and statistically fragile in baseline specifications, but becomes larger and statistically significant when interacted with presidential party. Still, the persistence of a positive association across interactions with presidential party suggests that alignment with government litigants confers some advantage in the promotion process, reinforcing theories that judicial behavior responds—consciously or unconsciously—to career incentives.

\subsection*{Selection versus Treatment in Elite Credentials}
Taken together, the results regarding elite credentials indicate that elite clerkships operate through more than simple selection on pre-existing ability. Although elite clerks publish more opinions early in their district court careers (Table \ref{tab:early_career_behavior}), they do not exhibit superior performance on quality measures such as reversal rates or opinion length. Elite law school graduates similarly show higher early productivity and citation impact but equivalent reversal rates.

The temporal dynamics provide sharper insight into the mechanism. The promotion advantage associated with elite clerkships is strongest in the earliest years on the bench and declines thereafter (Table~\ref{tab:clerkship_tenure_interactions}, Column 1). Clerkships appear to facilitate rapid early visibility—potentially through professional networks, sponsorship, or signaling within elite legal circles—allowing some judges to move quickly into appellate positions.

By contrast, elite law school credentials exhibit a stable association with promotion throughout judicial careers and show no tenure-dependent variation (Table~\ref{tab:clerkship_tenure_interactions}, Column 2). This difference suggests that law school prestige operates primarily through sustained reputational signaling, while clerkships provide access to networks that accelerate early-career promotion.

Importantly, both the clerkship and law school effect persists even within a subsample of judges with elite observable credentials at appointment, indicating that the premium cannot be fully attributed to baseline prestige. Instead, the evidence is more consistent with mechanisms operating after appointment—such as network access, mentorship, or institutional sponsorship—than with pure selection on ability.

While these findings do not establish causality, they meaningfully narrow the range of plausible explanations. Future work could further disentangle selection from treatment by exploiting institutional variation in access to elite clerkships, such as changes in the number of appellate judgeships over time. Even suggestive evidence along these lines would clarify whether clerkships operate primarily through network-based mechanisms rather than pre-existing differences in ability.

\subsection*{Limitations and Future Work}

Several limitations merit emphasis. Although the hazard framework improves interpretability relative to pooled models, the analysis remains observational. Unobserved factors such as informal networks, strategic retirement decisions, or behind-the-scenes lobbying may bias estimates. The inability to fully model senior status transitions may also affect age and tenure effects. Finally, while party-level heterogeneity is informative, it may mask meaningful variation across individual administrations.

Future research could extend this framework by incorporating administration-specific variation, Senate control, or explicit modeling of senior status decisions. More broadly, the approach developed here can be applied to other elite bureaucracies to study how political principals balance competence and loyalty in career advancement.

\section*{Conclusions} 
%%%%%%%%%%%%%%%%%%%%%%%%%%%%%%%%%%%%%%%%%%%%%%%%%%%%%%%%%%%%%%%%%%%%%%%%%%%%%%%%%%%

This paper reexamines judicial promotion through a discrete-time hazard framework that models promotion as a year-by-year process conditional on continued eligibility. This approach yields more stable estimates and clearer interpretation than pooled promotion models, allowing us to distinguish how political alignment, performance, and credentials shape promotion hazard over time. We find that promotions are strongly political: copartisanship with the sitting president substantially increases promotion probability. At the same time, performance signals such as productivity, reversal rates and citation centrality matter. Traditional credentials like ABA ratings and citation-based influence play a limited or contingent role, while elite credentials like law school and clerkship prestige do matter. Overall, judicial promotions reflect shifting trade-offs between ideology and competence that depend on the governing coalition, underscoring the political nature of judicial career advancement.

%%%%%%%%%%%%%%%%%%%%%%%%%%%%%%%%%%%%%%%%%%%%%%%%%%%%%%%%%%%%%%%%%%%%%%%%%%%%%%%%%%%
%%%%%%%%%%%%%%%%%%%%%%%%%%%%%%%%%%%%%%%%%%%%%%%%%%%%%%%%%%%%%%%%%%%%%%%%%%%%%%%%%%%
% References. 
%%%%%%%%%%%%%%%%%%%%%%%%%%%%%%%%%%%%%%%%%%%%%%%%%%%%%%%%%%%%%%%%%%%%%%%%%%%%%%%%%%%
%%%%%%%%%%%%%%%%%%%%%%%%%%%%%%%%%%%%%%%%%%%%%%%%%%%%%%%%%%%%%%%%%%%%%%%%%%%%%%%%%%%

%%%%%%%%%%%%%%%%%%%%%%%%%%%%%%%%%%%%%%%%%%%%%%%%%%%%%%%%%%%%%%%%%%%%%%%%%%%%%%%%%%%
%%%%%%%%%%%%%%%%%%%%%%%%%%%%%%%%%%%%%%%%%%%%%%%%%%%%%%%%%%%%%%%%%%%%%%%%%%%%%%%%%%%

%%%%%%%%%%%%%%%%%%%%%%%%%%%%%%%%%%%%%%%%%%%%%%%%%%%%%%%%%%%%%%%%%%%%%%%%%%%%%%%%%%%
%%%%%%%%%%%%%%%%%%%%%%%%%%%%%%%%%%%%%%%%%%%%%%%%%%%%%%%%%%%%%%%%%%%%%%%%%%%%%%%%%%%
\section*{Code Availability Statement} 
%%%%%%%%%%%%%%%%%%%%%%%%%%%%%%%%%%%%%%%%%%%%%%%%%%%%%%%%%%%%%%%%%%%%%%%%%%%%%%%%%%%
%%%%%%%%%%%%%%%%%%%%%%%%%%%%%%%%%%%%%%%%%%%%%%%%%%%%%%%%%%%%%%%%%%%%%%%%%%%%%%%%%%%

Code needed to replicate this study, in Python 3.7.6, is available at \url{https://github.com/ilyadavidson/judge-promotion-replication}.

%%%%%%%%%%%%%%%%%%%%%%%%%%%%%%%%%%%%%%%%%%%%%%%%%%%%%%%%%%%%%%%%%%%%%%%%%%%%%%%%%%%
%%%%%%%%%%%%%%%%%%%%%%%%%%%%%%%%%%%%%%%%%%%%%%%%%%%%%%%%%%%%%%%%%%%%%%%%%%%%%%%%%%%

{
\bibliographystyle{unsrt}
\bibliography{references.bib}
}

%%%%%%%%%%%%%%%%%%%%%%%%%%%%%%%%%%%%%%%%%%%%%%%%%%%%%%%%%%%%%%%%%%%%%%%%%%%%%%

%%%%%%%%%%%%%%%%%%%%%%%%%%%%%%%%%%%%%%%%%%%%%%%%%%%%%%%%%%%%%%%%%%%%%%%%%%%%%%
\section*{Competing Interest}
The authors declare no competing interests.

%%%%%%%%%%%%%%%%%%%%%%%%%%%%%%%%%%%%%%%%%%%%%%%%%%%%%%%%%%%%%%%%%%%%%%%%%%%%%%%%%%%

% appendix
%%%%%%%%%%%%%%%%%%%%%%%%%%%%%%%%%%%%%%%%%%%%%%%%%%%%%%%%%%%%%%%%%%%%%%%%%%%%%%%%%%%
%%%%%%%%%%%%%%%%%%%%%%%%%%%%%%%%%%%%%%%%%%%%%%%%%%%%%%%%%%%%%%%%%%%%%%%%%%%%%%%%%%%
\onecolumn
\newpage
\clearpage
\pagenumbering{arabic}
\setcounter{page}{1}
\appendix
\begin{center}
    {\fontsize{18}{22}\selectfont \bfseries SI Appendix for} \\
    \vspace{2em}
    {\fontsize{14}{22}\selectfont \bfseries What drives judge promotions?}
\end{center}

\label{sec:appendix}
\counterwithin{figure}{section} % make figure numbering dependent on the section
\counterwithin{table}{section}  % make table numbering dependent on the section
\setcounter{equation}{0}
\makeatletter
\makeatother

%%%%%%%%%%%%%%%%%%%%%%%%%%%%%%%%%%%%%%%%%%%%%%%%%%%%%%%%%%%%%%%%%%%%%%%%%%%%%%%%%%%
%%%%%%%%%%%%%%%%%%%%%%%%%%%%%%%%%%%%%%%%%%%%%%%%%%%%%%%%%%%%%%%%%%%%%%%%%%%%%%%%%%%

\section{}
\subsection{Judge–Year Panel Construction}
We construct a judge–year panel of federal district court judges observed annually from initial appointment until promotion to the appellate bench, retirement, death, or the end of the sample period. The panel spans the period from 1930 to the present and contains 36,194 judge–year observations drawn from 2,588 judges.

Each observation represents a judge-year in which the judge is eligible for promotion. Judges exit the observation set upon promotion. Judges who are never promoted are right-censored at the end of their observed careers.

\subsection{Tables and figures}
\begin{table}[htbp]
\centering
\caption{Sample Composition}
\label{tab:composition}
\setlength{\tabcolsep}{4pt}
\begin{tabular}{l l c c}
\hline
\hline
Variable & Category & Count & Percent \\
\hline
Era & 1930s & 454 & 1.25 \\
 & 1940s & 1{,}248 & 3.45 \\
 & 1950s & 1{,}722 & 4.76 \\
 & 1960s & 2{,}628 & 7.26 \\
 & 1970s & 3{,}780 & 10.44 \\
 & 1980s & 5{,}120 & 14.15 \\
 & 1990s & 5{,}874 & 16.23 \\
 & 2000s & 6{,}238 & 17.23 \\
 & 2010s & 5{,}729 & 15.83 \\
 & 2020s & 3{,}401 & 9.40 \\
\hline
Gender & Male & 30{,}248 & 83.57 \\
 & Female & 5{,}946 & 16.43 \\
\hline
Ethnicity & White & 30{,}747 & 84.95 \\
 & Black & 2{,}959 & 8.18 \\
 & Hispanic & 1{,}955 & 5.40 \\
 & Asian & 533 & 1.47 \\
\hline
Presidential party & Republican & 18{,}481 & 51.06 \\
 & Democratic & 17{,}713 & 48.94 \\
\hline
\hline
\end{tabular}
\end{table}

\subsection{Outcome Variable: Judicial Promotion}
The outcome of interest is promotion from a U.S. District Court to a U.S. Court of Appeals. We code a binary indicator equal to one in the year in which a judge is promoted and zero otherwise. Promotion events are rare, occurring in approximately 0.64 percent of judge–years.

\subsection{Discrete-Time Hazard Model}
We model promotion using a discrete-time hazard framework estimated via logistic regression. The model estimates the conditional probability that judge i is promoted in year t, given that the judge has not been promoted prior to year t. Formally,

\begin{equation}
h_{it}
= \Lambda(X_{it}\beta),
\end{equation}

where

\begin{equation}
h_{it} = \Pr(\text{Promotion}_{it} = 1 \mid \text{No promotion before } t)
\end{equation}

This framework allows promotion hazard to depend on both time-varying and time-invariant characteristics, accommodates right-censoring, and avoids the selection bias inherent in pooled or cross-sectional specifications that condition on eventual promotion.

Standard errors are clustered at the judge level to account for serial correlation within judicial careers.

\subsection{Baseline Hazard Specification}
To flexibly capture baseline promotion hazard over the judicial life cycle, we include categorical tenure bins (0–2, 3–5, 6–10, 11–15, 16–20, and 20+ years since district court appointment) and a quadratic specification for judicial age. These controls capture the well-documented “window of promotion” pattern in which promotion probabilities rise early in a judge’s career and decline at later stages.

We further include decade-of-appointment fixed effects to absorb cohort-specific differences in promotion opportunities and circuit fixed effects to control for persistent institutional differences across jurisdictions.

Unlike the continuous-time Cox model, which leaves the baseline hazard unspecified and therefore does not include an intercept, the discrete-time hazard model parameterizes baseline promotion hazard directly through time indicators. In our specification, tenure bins capture the shape of the baseline hazard, while the intercept represents the promotion hazard in the reference tenure category. Including an intercept is therefore necessary to anchor the baseline hazard and to interpret tenure coefficients as deviations from this reference period.

\subsection{Base Features}
Judicial age is centered at its sample mean prior to estimation. Centering improves numerical stability in models that include both linear and quadratic age terms by reducing collinearity between age and age squared. It also facilitates interpretation: the intercept and tenure effects correspond to promotion probabilities at the mean age, rather than at an arbitrary or meaningless value such as age zero. Centering does not affect fitted values or substantive conclusions, but ensures that age effects are interpreted as deviations around a typical career stage rather than absolute age levels.

\subsection{Political Alignment Variables}
\label{app:politicalalignment}
Political alignment is measured using two indicators: an indicator for whether the judge and the sitting president share partisan affiliation (copartisanship), and an indicator for the party of the sitting president.

Judicial elevation to the federal courts of appeals formally requires both presidential nomination and Senate confirmation \cite{uscourts_federal_judges_faq}. As a result, promotion outcomes reflect a joint institutional process rather than unilateral executive choice. Our analysis focuses exclusively on federal judges who were first appointed to the U.S. district courts and subsequently promoted to the courts of appeals. Judges appointed directly to the appellate bench without prior district court service are excluded due to data limitations, consistent with prior research on judicial promotion \cite{PosnerEpsteinLandes2013BehaviorFederalJudges}.

Because all judges in the hazard set have already undergone Senate confirmation at the time of their initial district court appointment, the Senate’s role in subsequent promotion decisions primarily functions as a feasibility constraint rather than as an active selector. Historical evidence supports this characterization: between 1939 and 2014, fewer than one percent of judicial nominations to the district or appellate courts were rejected by the Senate \cite{crs_R40470}. This low rejection rate likely reflects the fact that promoted judges have already passed prior confirmation and thus constitute a pre-screened pool of candidates acceptable to the legislative branch.

Accordingly, the estimated hazard of promotion should be interpreted as the probability that a judge is elevated in a given year conditional on remaining eligible and acceptable to both the nominating administration and the confirming Senate. Variables capturing partisan alignment—such as copartisanship between the judge and the sitting president—therefore proxy for executive selection within a legislatively constrained set of candidates, rather than implying exclusive presidential control over promotion outcomes.

\subsection{Elite Credentials}
Elite credentials are captured using indicators for attendance at a highly ranked law school and completion of a prestigious judicial clerkship. These measures proxy for elite signaling, professional networks, and reputational capital that may influence promotion decisions independently of observable judicial behavior. We classify law schools as elite based on the 1987 U.S. News \& World Report rankings, the earliest year for which comprehensive rankings are available. This approach follows Choi and Gulati (2004) and ensures the classification predates the careers of most judges in our sample. The top five in 1987 were Harvard, Yale, Columbia, Michigan, and Stanford.

\subsection{Judicial Performance and Behavior Measures}
\label{app:judicialperformance}
We incorporate multiple measures of judicial performance and behavior, including appellate reversal rates, citation impact, publication intensity, and opinion length.

For the reversal rate, we obtained appellate outcomes by matching district court opinions to subsequent circuit court decisions using case citations from CourtListener. We code a case as "reversed" if the appellate court's disposition is anything other than affirmance. This includes: reversed (in whole or in part), vacated, remanded for reconsideration, modified or amended. Cases that are affirmed or where no appeal was filed are coded as non-reversed.

To reduce noise from small case counts, especially early in judicial careers, we apply empirical Bayes (EB) shrinkage to reversal rates. EB shrinkage pulls individual judge estimates toward the population mean, with the degree of shrinkage inversely proportional to the number of cases. This approach stabilizes estimates and ensures that observed effects reflect persistent judicial tendencies rather than idiosyncratic outcomes in a small number of cases. Formally, for judge \textit{j} with \textit{n} cases and \textit{r} reversals:

\begin{equation*}
    \text{EB Reversal Rate} =  \frac{(r + \alpha)}{(n + \alpha + \beta)}
\end{equation*}

where $\alpha$ and $\beta$ are estimated from the population distribution of reversal rates using maximum likelihood.

For each judge-year, we calculate the reversal rate using all cases decided by that judge from their appointment through year \textit{t}. This creates a time-varying measure that accumulates as judges gain experience and develop track records.

We construct a directed citation network using the complete corpus of federal case law from CourtListener and the Caselaw Access Project. The network contains 740,887 cases (nodes) and 5,183,294 citations (directed edges) spanning federal district courts, circuit courts, and the Supreme Court from 1880 to present. When Case A cites Case B in its opinion, we create a directed edge from A to B. This edge indicates that B's legal reasoning influenced A's decision. The network is directed because citation relationships are asymmetric: A cites B, but B does not necessarily cite A. For each case in the network, we calculate PageRank centrality using the standard PageRank algorithm (Page et al. 1999). PageRank assigns centrality scores based on the principle that a case is central if it is cited by other central cases. Formally:

\begin{equation*}
    \text{PR}(\text{case}_i) = \frac{(1-d)}{N} + d \times \sum_{j \in \text{In}(i)} \frac{\text{PR}(\text{case}_j)}{\text{out\_degree}(\text{case}_j)}
\end{equation*}

where the sum is over all cases j that cite case i, d is the damping factor (set to 0.85), and N is the total number of cases. We iterate until convergence (typically 30-50 iterations).

Citation networks exhibit power-law distributions with extreme outliers. A small number of landmark cases have centrality scores orders of magnitude higher than typical cases. To prevent these outliers from dominating the analysis, we winsorize centrality at the 1st and 99th percentiles before aggregating to the judge level. This means any centrality score below the 1st percentile is set to the 1st percentile value, and any score above the 99th percentile is set to the 99th percentile value. Without winsorization, the maximum centrality is 47.365 standard deviations above the mean; after winsorization, it is 2.452. This reduces the coefficient from $\beta$ = 0.071 (p = 0.006) to $\beta$ = 0.230 (p = 0.025), suggesting that judges with extraordinarily influential landmark cases were not promoted and suppressed the average effect.

For each judge-year observation, we calculate the average PageRank centrality across all published opinions authored by that judge from appointment through year t-1. This creates a cumulative, time-varying measure of a judge's influence within the legal citation network.

An alternative approach would construct a judge-to-judge citation network, where an edge from Judge i to Judge j exists if any case authored by Judge i cites any case authored by Judge j. We could then measure centrality at the judge level directly. We use the case-level approach because: (1) it captures heterogeneity in influence across a judge's opinions, (2) it aligns with the legal reality that individual cases, not judges, serve as precedent, and (3) it allows tracking how influence evolves as judges accumulate cases. Results using judge-level network centrality are substantively similar (available upon request).

We tested both raw citation counts (total number of citations to a judge's opinions) and network centrality as predictors. Raw citations showed no relationship with promotion ($\beta$ = -0.036, p = 0.677), while network centrality was significant ($\beta$ = 0.230, p = 0.025). This suggests presidents and senators reward judges whose cases are cited by other influential decisions (network position), not merely judges with high citation volumes.

Publication intensity is measured as the cumulative number of authored, published opinions divided by years of service, yielding an average annual publication rate, which is then log-transformed and standardized.Opinion length is measured as the mean of log-transformed word counts at the judge–year level, rather than the log of mean opinion length. This approach captures a judge’s typical writing style while limiting the influence of occasional unusually long opinions and reduces sensitivity to outliers.

To capture judicial behavior toward powerful litigants, we construct cumulative measures of support for the federal government and for large corporate parties in civil cases.

All performance and behavioral measures are standardized to have mean zero and unit variance to facilitate comparison of effect sizes across covariates.

\subsection{Party-Specific Promotion Criteria}
In the final specification, we allow key performance and behavioral measures to vary by presidential party through interaction terms. This specification tests whether the criteria used to evaluate judicial candidates differ systematically across partisan administrations, rather than assuming a single, time-invariant promotion rule.

\subsection{Mechanisms Behind Elite Clerkship Effects}
\begin{table}[htbp]
\centering
\small
\setlength{\tabcolsep}{0.5pt}
\caption{Early-Career Judicial Behavior and Elite Credentials}
\label{tab:early_career_behavior}
\begin{tabular}{lcccc}
\hline\hline
 & \multicolumn{2}{c}{Elite Law School (LS$_p$)} & \multicolumn{2}{c}{Elite Clerkship (C$_p$)} \\
Outcome & Coef. & S.E. & Coef. & S.E. \\
\hline
Publications & $0.103^{***}$ & $(0.023)$ & $0.066^{**}$ & $(0.025)$ \\
Opinion length (log) & $0.043^{*}$ & $(0.025)$ & $-0.034$ & $(0.026)$ \\
Reversal rate & $-0.036$ & $(0.027)$ & $-0.018$ & $(0.032)$ \\
Citation impact & $0.060^{**}$ & $(0.029)$ & $-0.015$ & $(0.029)$ \\
\hline\hline
\end{tabular}
\begin{flushleft}
\footnotesize
Notes: Table reports coefficients from OLS regressions estimated on judge-year observations restricted to the first five years following appointment to the district bench. Each outcome is regressed separately on elite law school attendance (LS$_p$) and elite clerkship experience (C$_p$), controlling for era, circuit, age, gender, ethnicity, and other baseline covariates. Standard errors are clustered at the judge level.  
$^{*} p<0.10$, $^{**} p<0.05$, $^{***} p<0.01$.
\end{flushleft}
\end{table}

\subsubsection{Within-Elite Subsample Robustness}
\begin{table}[htbp]
\centering
\small
\caption{Promotion Effects Within Elite Subsample}
\label{tab:within_elite_promotion}
\setlength{\tabcolsep}{1pt}
\begin{tabular}{l c c}
\hline\hline
Variable & Coefficient & Std. Error \\
\hline
Elite Law School (LS$_p$) & 0.359$^{**}$ & (0.175) \\
Elite Clerkship (C$_p$)  & 0.663$^{***}$ & (0.238) \\
\hline\hline
\end{tabular}
\begin{flushleft}
\footnotesize
Notes: Entries are log-odds coefficients from a discrete-time hazard model estimated on the subsample of judges with elite credentials at appointment. The model includes controls for era, circuit, age, tenure, demographics, political alignment, and judicial performance, with judge-level clustered standard errors.  
$^{*} p<0.05$, $^{**} p<0.01$, $^{***} p<0.001$.
\end{flushleft}
\end{table}
\newpage

\clearpage
\subsubsection{Interaction and Placebo Tests}
\begin{table}[htbp]
\centering
\small
\caption{Elite Clerkships, Tenure, and Promotion Hazards}
\label{tab:clerkship_tenure_interactions}
\setlength{\tabcolsep}{0.5pt}
\begin{tabular}{lcc}
\hline\hline
 & (1) & (2) \\
 & Clerkship $\times$ Tenure & Law School $\times$ Tenure (Placebo) \\
\hline
Elite credential (baseline) 
& $1.44^{***}$ & $0.32$ \\
& $(0.40)$ & $(0.40)$ \\[4pt]

$\times$ Tenure 3--5 years
& $-1.59^{**}$ & $-0.29$ \\
& $(0.62)$ & $(0.49)$ \\

$\times$ Tenure 6--10 years
& $-1.32^{*}$ & $0.17$ \\
& $(0.63)$ & $(0.51)$ \\

$\times$ Tenure 11--15 years
& $-0.72$ & $0.11$ \\
& $(0.57)$ & $(0.50)$ \\

$\times$ Tenure 16--20 years
& $-0.51$ & $-0.04$ \\
& $(0.66)$ & $(0.60)$ \\

$\times$ Tenure 20+ years
& $-0.96$ & $0.97$ \\
& $(1.33)$ & $(0.96)$ \\

\hline
Controls & Yes & Yes \\
Era FE & Yes & Yes \\
Circuit FE & Yes & Yes \\
Observations & 36,194 & 36,194 \\
Judges & 2,588 & 2,588 \\
\hline\hline
\end{tabular}

\begin{flushleft}
\footnotesize
Notes: Entries are log-odds coefficients from discrete-time hazard models.
The omitted tenure category is 0--2 years. Column (1) interacts elite
clerkship status with tenure bins; Column (2) reports a placebo interaction
using elite law school credentials. Standard errors clustered at the judge level.
$^{*}p<0.05$, $^{**}p<0.01$, $^{***}p<0.001$.
\end{flushleft}
\end{table}

\subsection{IV Analysis of Reversal Rates}
\label{app:ivanalysis}
To construct the instrument, for each district judge $d$ we collect all appealed cases involving that judge. For each such case $c$, we recompute the panel's reversal propensity using appellate-judge reversal rates that exclude cases involving judge $d$. That is, for each appellate judge $a$ on panel $P_c$,

\[
r_{a,-d} = \frac{\# \text{ reversed cases heard by } a \text{ not involving } d}{\# \text{ cases heard by } a \text{ not involving } d}.
\]

Using these leave-out propensities, we construct three alternative measures of case-level panel harshness:

\[
p^{\text{mean}}_{c,-d} = \frac{1}{|P_c|}\sum_{a \in P_c} r_{a,-d}, \quad
p^{\text{min}}_{c,-d} = \min_{a \in P_c} r_{a,-d}, \quad
p^{\text{max}}_{c,-d} = \max_{a \in P_c} r_{a,-d}.
\]

These correspond to different assumptions about how individual judges' tendencies aggregate within a panel: the mean treats all panel members equally, the minimum captures the influence of the most lenient judge, and the maximum captures the most reversal-prone judge. For each district judge $d$, we aggregate across all appealed cases $\mathcal{C}_j$ to obtain three
judge-level instruments \footnote{The instrument $Z_j$ is computed as a time-invariant judge-level average across all appealed cases in a judge's career.
  This choice rests on the assumption that a judge's exposure to panel harshness reflects a stable feature of their appellate
  environment, a form of latent harshness, rather than year-to-year variation. Under this assumption, the instrument
  captures systematic differences in reversal exposure across judges attributable to the panels they faced, rather than
  idiosyncratic case-level noise. A time-varying construction would reduce statistical power substantially given the rarity of promotion events.}:

\[
Z^{\text{mean}}_d = \frac{1}{N_j}\sum_{c \in \mathcal{C}_d} p^{\text{mean}}_{c,-d}, \quad
Z^{\text{min}}_d = \frac{1}{N_j}\sum_{c \in \mathcal{C}_d} p^{\text{min}}_{c,-d}, \quad
Z^{\text{max}}_d = \frac{1}{N_j}\sum_{c \in \mathcal{C}_d} p^{\text{max}}_{c,-d}.
\]

We require at least 3 appealed cases per district judge, yielding the IV subsample described in Table ~\ref{tab:ivsubsample}.

\begin{table}[H]
\centering
\caption{Sample construction and instrument coverage}
\label{tab:ivsubsample}
\begin{tabular}{lcc}
\toprule
Quantity & Value  \\
\midrule
Original hazard panel (judge-years) & 36,194  \\
Judge-years with non-missing instrument & 27,187  \\
Original promotions & 232 \\
Promotions in the IV sample & 116 \\ 
Judges with instrument & 1,916 \\
\bottomrule
\end{tabular}
\end{table}

\begin{figure}[H]
  \centering
  \includegraphics[width=\textwidth]{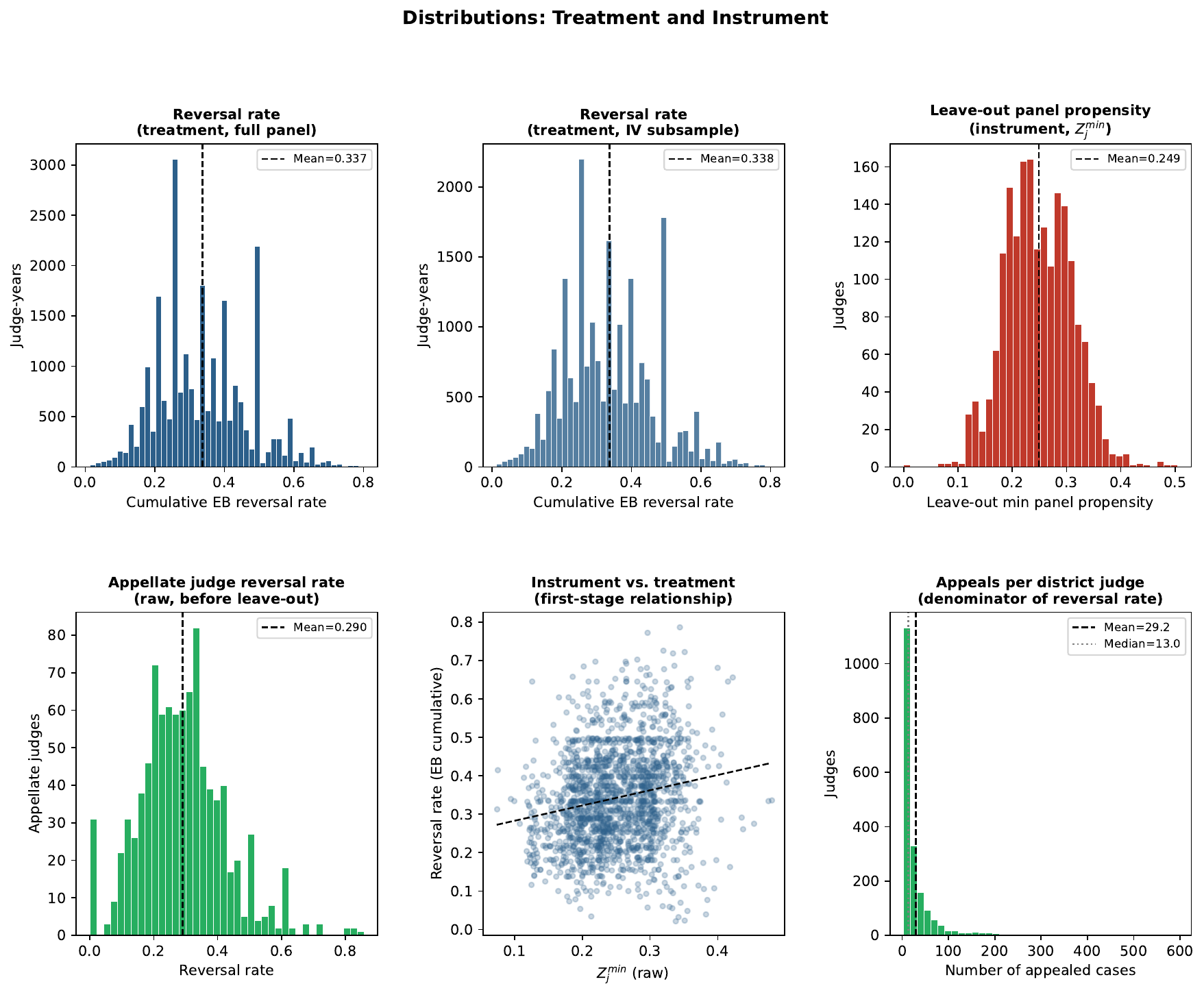}
  \caption{Distributions of treatment and instrument variables. \textit{Top row, left to right}: cumulative EB reversal
rate in the full hazard panel; cumulative EB reversal rate in the IV subsample; leave-out min panel propensity
($Z^{\text{min}}$). \textit{Bottom row}: raw appellate judge reversal rates (before leave-out); first-stage scatter plot of
instrument vs.\ treatment; violin plots of all three instrument variants ($Z^{\text{min}}$, $Z^{\text{mean}}$,
$Z^{\text{max}}$).}
  \label{fig:iv_distributions}
\end{figure}

Figure~\ref{fig:iv_distributions} displays the distributions of the key variables underlying the IV analysis. The top row
shows the cumulative empirical Bayes reversal rate for the full hazard panel and the IV subsample, confirming that the
subsample is not systematically different from the full sample, alongside the distribution of the preferred leave-out panel
propensity instrument ($Z^{\text{min}}$). The bottom row shows the raw reversal rates of individual appellate judges, the
first-stage scatter plot visualizing the relationship between instrument and treatment, and the bottom-right panel shows the distribution of appealed cases per district judge.

Table ~\ref{tab:iv_results} reports first- and second-stage estimates for all three instrument variants. Across specifications, the
minimum-based instrument ($Z^{\text{min}}$) achieves the strongest first-stage fit (F = 7.04), suggesting that the most
lenient judge on the panel is the primary driver of a district judge's reversal exposure. The mean- and maximum-based
instruments yield substantially weaker first stages (F = 3.32 and F = 0.72, respectively), rendering their second-stage
estimates unreliable. We therefore focus on the minimum-based instrument as our preferred specification, a choice further
supported by the balance tests below.

\begin{table}[H]\centering
\caption{IV Estimates: Panel Harshness Exposure and Promotion}
\label{tab:iv_results}
\begin{tabular}{lccc}
\toprule
 & (1) Min IV & (2) Mean IV & (3) Max IV \\
\midrule

\textit{First stage} & & & \\
Leave-out panel propensity 
& 0.0736*** & 0.9176* & 0.0261 \\
& (0.0277) & (0.5039) & (0.0308) \\
F-statistic 
& 7.04 & 3.32 & 0.72 \\

\addlinespace
\textit{Second stage (IV)} & & & \\
Reversal rate (instrumented)
& 2.4897 & 8.4135*** & 22.3668*** \\
& (2.3749) & (2.4994) & (4.4863) \\

Control function residual
& -2.6574 & -8.5815*** & -22.5343*** \\
& (2.3739) & (2.4891) & (4.4777) \\

\midrule
Observations & 26,707 & 26,707 & 26,707 \\
R-squared (first stage) & 0.023 & 0.022 & 0.021 \\
\bottomrule

\multicolumn{4}{p{0.85\textwidth}}{\footnotesize
Notes: The table reports first-stage and second-stage estimates using alternative constructions of the leave-out panel propensity instrument (minimum, mean, and maximum across appellate panels). All models include circuit fixed effects and standard controls. Standard errors are clustered at the judge level and reported in parentheses. The control-function approach is used for the nonlinear second stage. The F-statistic refers to the excluded instrument in the first stage. \
*** $p<0.01$, ** $p<0.05$, * $p<0.10$.
}
\end{tabular}
\end{table}

\subsection*{Balance tests}
A valid instrument must be independent of pre-determined judge characteristics. Table ~\ref{tab:iv_balance} reports balance tests in which
each instrument is regressed on elite credentials and political alignment. The minimum-based instrument passes these checks:
none of the coefficients on $LS_p$, $C_p$, or copartisanship reach conventional significance. The mean- and maximum-based
instruments fail on multiple dimensions, with significant correlations with both elite law school status and clerkship
credentials. This confirms that $Z^{\text{min}}$ is the only specification satisfying both relevance and exogeneity, and
further justifies its use as the preferred instrument.

\begin{table}[H]\centering
\caption{Balance / Exogeneity Tests for Alternative Instruments}
\label{tab:iv_balance}
\begin{tabular}{lccc}
\toprule
 & (1) Min IV & (2) Mean IV & (3) Max IV \\
\midrule
\textit{Elite credentials and political alignment} \\
LS\_p        & 0.0853*   & 0.0922**  & 0.1000** \\
             & (0.054)   & (0.027)   & (0.018)  \\
C\_p         & -0.0274   & -0.1369*** & -0.2223*** \\
             & (0.597)   & (0.005)   & (0.000)  \\
copartisan   & -0.0106   & 0.0639    & 0.1132   \\
             & (0.917)   & (0.499)   & (0.239)  \\
\midrule
Observations & \multicolumn{3}{c}{\textit{Judge-level sample}} \\
\bottomrule
\end{tabular}

\vspace{0.3cm}
\footnotesize
\textit{Notes:} The table reports balance tests where each instrument (minimum, mean, and maximum leave-out panel propensity) is regressed on predetermined judge characteristics. Reported are coefficients with p-values in parentheses. *** $p<0.01$, ** $p<0.05$, * $p<0.10$.
\end{table}

\end{document}